\documentclass[prb,cha,twocolumn,nobibnotes,preprintnumbers,showpacs, amsmath,amssymb]{revtex4-1}
\usepackage{bm}
\usepackage{hyperref}
\hypersetup{colorlinks=true,linkcolor=black,citecolor=black,urlcolor=black}
\usepackage{amsmath}
\usepackage{amssymb}
\usepackage{amsthm}
\usepackage{amsfonts}
\usepackage{enumerate}
\usepackage{latexsym}
\usepackage{ifpdf}
\usepackage{graphicx}
\usepackage{makeidx}
\usepackage{multirow}

\expandafter\ifx\csname package@font\endcsname\relax\else
\expandafter\expandafter
\expandafter\usepackage
\expandafter\expandafter
\expandafter{\csname package@font\endcsname}
\fi
\usepackage[section]{placeins}
\usepackage{times}
\usepackage[normalem]{ulem}

\usepackage{epstopdf}

\begin{document}

\title{Unique quantum metallic state in the titanium sesquioxide heterointerface superconductor}

\author{Guanqun Zhang,$^{1}$ Yixin Liu,$^{2,3}$ Zhongfeng Ning,$^{1}$ Guoan Li,$^{4}$ Jinghui Wang,$^{5}$ Yueshen Wu,$^{5}$ Lijie Wang,$^{1}$ Huanyi Xue,$^{1}$ Chunlei Gao,$^{1,6}$ Zhenghua An,$^{1,6}$ Jun Li,$^{5}$ Jie Shen,$^{4,7}$ Gang Mu,$^{2,3,*}$ Yan Chen,$^{1,*}$ and Wei Li$^{1,*}$}

\affiliation
{$^1$State Key Laboratory of Surface Physics and Department of Physics, Fudan University, Shanghai 200433, China\\
 $^2$National Key Laboratory of Materials for Integrated Circuits, Shanghai Institute of Microsystem and Information Technology, Chinese Academy of Sciences, Shanghai 200050, China\\
 $^3$University of Chinese Academy of Sciences, Beijing 100049, China\\
 $^4$Beijing National Laboratory for Condensed Matter Physics and Institute of Physics, Chinese Academy of Sciences, Beijing 100190, China\\
 $^5$ShanghaiTech Laboratory for Topological Physics, School of Physical Science and Technology, ShanghaiTech University, Shanghai 201210, China\\
 $^6$Institute for Nanoelectronic Devices and Quantum Computing, Fudan University, Shanghai 200433, China\\
 $^7$Songshan Lake Materials Laboratory, Dongguan 523808, China
 }

\date{\today}

\begin{abstract}

\textbf{ABSTRACT:}  The emergence of quantum metallic state marked by a saturating finite electrical resistance in the zero-temperature limit in a variety of two-dimensional superconductors injects an exciting momentum to the realm of heterostructure superconductivity. Despite much research efforts over last few decades, there is not yet a general consensus on the nature of this unexpected quantum metal. Here, we report the observation of a unique quantum metallic state within the hallmark of Bose-metal in the titanium sesquioxide heterointerface superconductor Ti$_2$O$_3$/GaN. Remarkably, the quantum bosonic metallic state continuously tuned by a magnetic field in the vicinity of the two-dimensional superconductivity-metal transition persists in the normal phase, indicating the existence of composite bosons formed by electron Cooper pairs even in the normal phase. Our findings provide a distinct evidence for electron pairing in the normal phase of heterointerface superconductors, and shed fresh light on the pairing nature underlying heterointerface superconductivity.

\textbf{KEYWORDS:}  \textit{heterointerface}, \textit{two dimensions}, \textit{superconductivity}, \textit{quantum metal}, \textit{electron Cooper pairs}
\end{abstract}

\maketitle

\section*{INTRODUCTION}

\begin{figure*}
\centering
\includegraphics[bb=35 442 580 700,width=\textwidth]{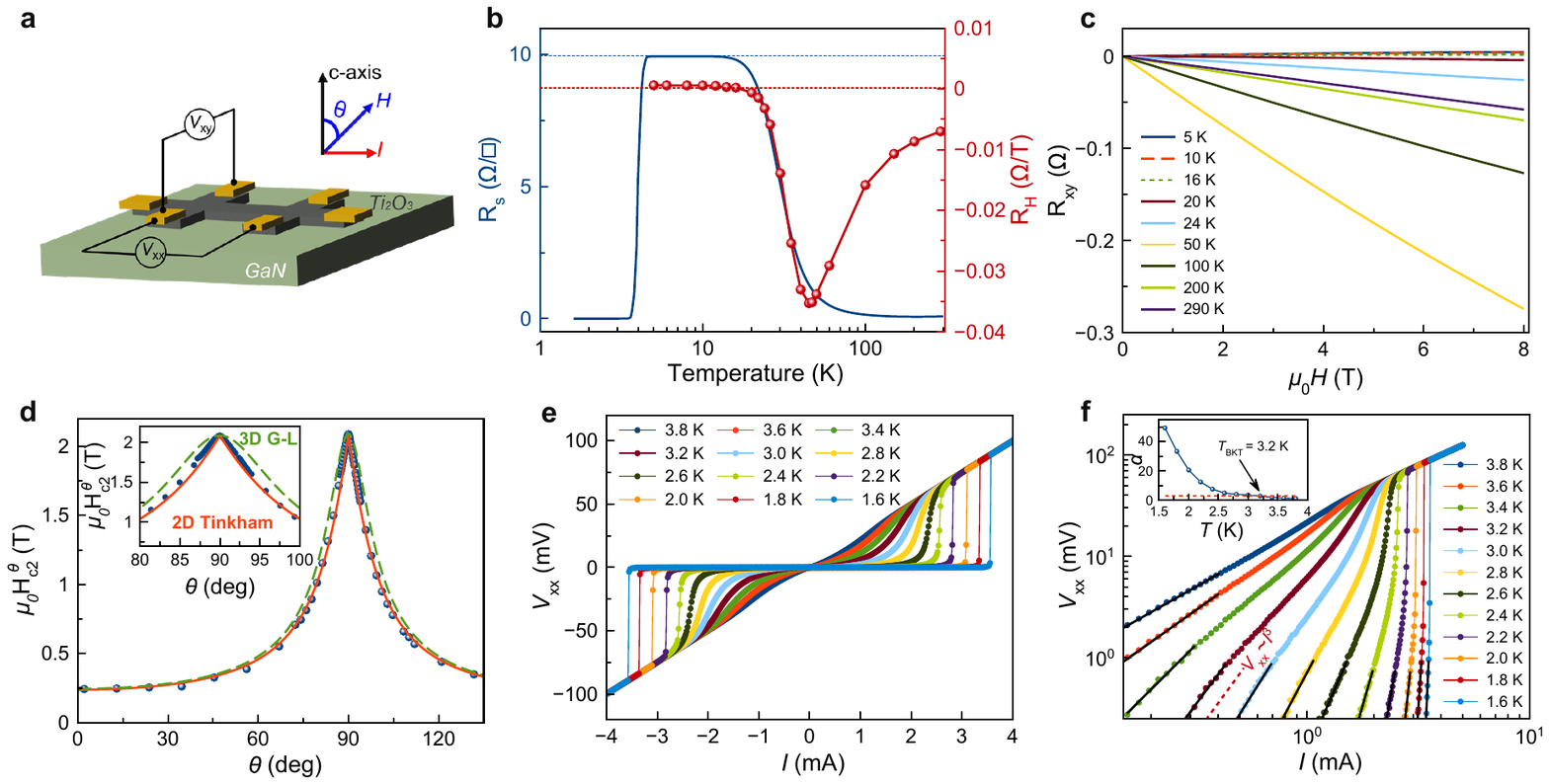}
\caption{Two-dimensional superconductivity of Ti$_2$O$_3$/GaN. (a) Schematic structure of the Hall bar device on Ti$_2$O$_3$/GaN. (b) Longitudinal electrical resistance $\mathrm{R}_{\mathrm{s}}$ and the corresponding transverse Hall coefficient $\mathrm{R}_{\mathrm{H}}$(=$d\mathrm{R}_{\mathrm{xy}}/d\mu_0H$) as a function of temperature. (c) Transverse Hall resistance $\mathrm{R}_{\mathrm{xy}}$ as a function of an applied out-of-plane field with various temperatures ranging from 5 to 300 K. (d) Out-of-plane polar angular $\theta$ dependence of the upper critical field $\mu_0\mathrm{H}^{\theta}_{\mathrm{c2}}$. The inset is a magnification of the region around $\theta=90^{\circ}$. The solid line represents the fit obtained by the two-dimensional (2D) Tinkham model, and the dashed line represents the fit by the three-dimensional (3D) anisotropic Ginzburg-Landau (G-L) model. (e) Temperature-dependent $I$-$V_{\mathrm{xx}}$ measurements. (f) Corresponding logarithmic scale representation of (e). The red dashed line denotes the $V_{\mathrm{xx}}\sim I^3$ dependence. The inset shows the extracted power-law fitting exponent $\alpha$ as a function of the temperature. The BKT transition temperature $T_{\mathrm{BKT}}= 3.2$ K is defined by $\alpha=3$.}
\label{fig1}
\end{figure*}

The discovery of superconductivity at two-dimensional heterointerfaces has opened up intriguing avenues in frontiers of nanoscience and nanotechnology, which sparks a surge of research attention to unveil the underlying rich physical properties and provide an exciting opportunity for the development of novel mesoscopic superconducting circuits~\cite{Ref1,Ref2,Ref3,Ref4}. A variety of emergent appealing quantum behaviors has been observed in SrTiO$_3$ and KTaO$_3$ heterointerface superconductors~\cite{Ref5,Ref6}, including strong Rashba-like spin-orbit coupling~\cite{Ref7,Ref8,Ref9,Ref10,Hua2022}, interfacial ferromagnetism~\cite{Ref11,Ref12,Ref13,Ref14,ZNing2023}, and gate-tunable superconductivity~\cite{Ref15,Ref16,Ref17,Ref18,Ref19}. Remarkably, the superconducting to insulating quantum phase transition can be tuned continuously by the application of a gate voltage without introducing unintentional disorders. In the intermediate regime between the superconducting and insulating phases, a quantum metallic state with saturating finite electrical resistance exists and is stabilized in the ultralow temperature limit~\cite{Ref16,Ref19}, challenging our fundamental understanding of the conventional Fermi liquid theory of electron fluids because of the fact that electrons usually do not form a conventional metal in two dimensions prohibited by the Anderson localization~\cite{Anderson1958,Abrahams1979}. Although qualitatively similar behavior has also been extensively observed in disordered thin films~\cite{Paalanen,Ephron,Yazdani,Goldman2002,WLiu,Goldman2012,Chervenak2000,Mason}, exfoliated two-dimensional superconductors under applied magnetic fields~\cite{Nojima2015,Pasupathy2016,TWang}, and artificially patterned superconducting islands~\cite{Eley,YLi2019}, the existence and origin of such an unexpected quantum metallic state remain elusive~\cite{Kapitulnik,Fisher,Phillips}, which have hindered experimental progress, motivating the search for a more favorable heterointerface superconductor that hosts the intriguing quantum metallic phase with a distinctive origin.

Very recently, using state-of-the-art heterostructure design strategy, an impressive quantum metallic-like state has been experimentally observed at a finite high temperature at the superconducting heterointerface between the Mott insulator Ti$_2$O$_3$ and the polar semiconductor GaN as highlighted in our notable study~\cite{LJWang2022}. Specifically, the quantum metallic-like state is manifested at the verge of interfacial superconductivity, where the electrical resistance in the normal phase shows a wide temperature-independent plateau. In addition, the transverse Hall resistance is simultaneously vanishing, unveiling the existence of electron-hole symmetry inherent to the quantum metallic state within the hallmark of the Bose-metal scenario~\cite{Breznay,ZChen2021}. Notably, this intriguing quantum bosonic metallic state is experimentally characterized by the temperature-independent plateau of longitudinal electrical resistance and simultaneously vanishing Hall resistance~\cite{LJWang2022}. These results are in stark contrast to the basic picture of the emergence of the quantum metallic phase in the ultralow temperature limit~\cite{Kapitulnik,Klapwijk2020}. Thus, it is pivotal to clarify experimentally the deep connection between the quantum bosonic metallic state located in the normal phase and that of appearing at the superconducting to insulating phase transition tuned by fields at ultralow temperatures, offering innovative perspective on the underlying nature of the quantum metallic state.

In this work, we use a magnetic field to systemically tune the superconducting electronic behaviors at the heterointerface of Ti$_2$O$_3$/GaN through the superconductivity-metal transition. At first, we briefly revisit the two-dimensional superconductivity and the emergent quantum bosonic metal in the normal phase in Ti$_2$O$_3$/GaN. Then, we exploit the joint observations of saturating resistance and vanishing Hall resistance at the ultralow temperatures to identify the field driving superconductivity into the quantum bosonic metallic state. Finally, the quantum bosonic metallic state located in the normal phase is uniquely linked to the one neighbouring two-dimensional superconductivity-metal transition tuned by fields, pointing to a common origin of Bose-metal. These intriguing results thus provide a distinct evidence for the existence of composite bosons formed by electron Cooper pairs persisting in the normal phase, which is essential for understanding the underlying key ingredient of pairing mechanism in heterointerface superconductivity. 

\section*{RESULTS AND DISCUSSION}

An antiferromagnetic Mott insulating Ti$_2$O$_3$ thin film with a narrow-band gap of 0.1 eV~\cite{Morin,Iguchi} is grown by pulsed laser deposition (PLD) on top of a (0001)-oriented GaN substrate with an N-polar terminated face. The detailed sample growth and characterizations are described in Experimental Section and in our previous study~\cite{LJWang2022}. The thickness of the Ti$_2$O$_3$ thin film is about 90 nm, as determined by an atomic force microscope (AFM). Before proceeding with the electrical transport measurements, the active area of the Ti$_2$O$_3$ thin film device is patterned into a Hall bar configuration (see Figure~\ref{fig1}a and Figure S1). All transport measurements in this study are performed on the same device with an estimated electron-like charge carrier density of 8.67$\times$10$^{16}$ cm$^{-2}$ at a room temperature of 290 K (see also Figure~\ref{fig1}b,c). To avoid misalignment effects of the transverse contact pads in the Hall bar device, the mixing between longitudinal electrical resistance $\mathrm{R}_{\mathrm{s}}$ and transverse Hall resistance $\mathrm{R}_{\mathrm{xy}}$ have been minimized by reporting the field symmetrized longitudinal electrical resistances and anti-symmetrized transverse Hall resistances, respectively.

Figure~\ref{fig1}b shows $\mathrm{R}_{\mathrm{s}}$ as a function of temperature in the range from 1.5 to 300 K at zero magnetic field in Ti$_2$O$_3$/GaN. In the low temperature regime, we observe the conspicuous signal of superconducting transition to zero-resistance state, measured to the limit of our instrument resolution. The critical temperature is $T_c=3.98$ K, defined as the temperature where the resistance is at the midpoint of the normal phase value at 4.5 K. To further clarify the superconducting properties at the heterointerface of Ti$_2$O$_3$/GaN, we have characterized the sample by the following two measurements. First, the out-of-plane polar angular $\theta$ dependence of the upper critical field $\mu_0\mathrm{H}^{\theta}_{\mathrm{c2}}$ is shown in Figure~\ref{fig1}d. This is defined as the magnetic field at the midpoint of the electrical resistance transition with $\theta$ being the angle between the magnetic field and the perpendicular direction to the surface of Ti$_2$O$_3$/GaN (Figure~\ref{fig1}a, inset). When the magnetic field is parallel to the film ($\theta = 90^{\circ}$), $\mu_0\mathrm{H}^{\theta=90^{\circ}}_{\mathrm{c2}}$ is apparently much higher than that in the perpendicular field, $\mu_0\mathrm{H}^{\theta=0^{\circ}}_{\mathrm{c2}}$. The strong anisotropy in the observed superconducting critical fields illustrates the two-dimensional superconducting feature in Ti$_2$O$_3$/GaN (see also Figure S2). Moreover, at around $\theta = 90^{\circ}$, a cusp-like peak can be clearly seen (Figure~\ref{fig1}d, inset) and is qualitatively distinct from the three-dimensional anisotropic Ginzburg-Landau model, but is well described by the two-dimensional Tinkham model~\cite{Tinkham}, as frequently observed in interfacial superconductivity~\cite{LJWang2022,GZhang,ZNing2023}, surface of doped SrTiO$_3$~\cite{KUeno2014,Kozuka}, and layered transition metal dichalcogenides~\cite{JMLu,DJiang}. Then, we measure the current-voltage ($I$-$V_{\mathrm{xx}}$) characteristics at various temperatures close to $T_c$ (see Figure~\ref{fig1}e), which are shown using log-log scale in Figure~\ref{fig1}f. A power-law dependence of $V_{\mathrm{xx}} \propto I^{\alpha}$ behavior can be clearly observed. The exponent $\alpha$ decreases monotonically with temperature (Figure~\ref{fig1}f, inset). Here, the Berezinskii-Kosterlitz-Thouless (BKT) transition temperature $T_{\mathrm{BKT}}$ defines the dissociation of vortex-antivortex pairs in two-dimensional superconductors obeying the universal scaling relation $V_{\mathrm{xx}} \sim I^3$~\cite{Kosterlitz,Beasley}. We thus determine $T_{\mathrm{BKT}} = 3.2$ K from where $\alpha =3$ interpolates, consistent with the $T_c$ as defined in Figure~\ref{fig1}b.

\begin{figure}
\centering
\includegraphics[bb=85 230 295 420,width=9.5cm,height=8cm]{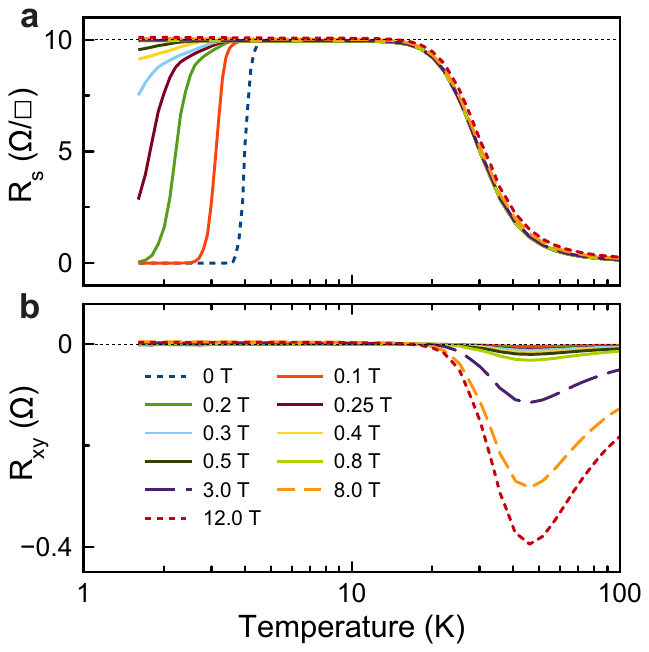}
\caption{Two-dimensional superconductivity-quantum bosonic metal transition driven by magnetic field in Ti$_2$O$_3$/GaN. (a) Longitudinal electrical resistance $\mathrm{R}_{\mathrm{s}}$ and (b) transverse Hall resistance $\mathrm{R}_{\mathrm{xy}}$ as a function of temperature in the presence of an out-of-plane magnetic field.}
\label{fig2}
\end{figure}

Remarkably, at the verge of superconducting phase in Ti$_2$O$_3$/GaN, we observe a pronounced temperature-independent $\mathrm{R}_{\mathrm{s}}$ in the normal phase that exhibits a plateau of finite resistance with a wide temperature range of 10 K (Figure~\ref{fig1}b and Figure S3). In the same range, the Hall resistance $\mathrm{R}_{\mathrm{xy}}$ is vanishing within the measurement resolution (see Figure~\ref{fig1}b,c) as a result of an inherent electron-hole symmetry~\cite{Breznay}. The temperature-independent plateau in the electrical resistance and the simultaneous vanishing of the Hall resistance are the distinct hallmark of a quantum metallic state within the Bose-metal scenario~\cite{Phillips,Breznay}. These intriguing experimental results shown in Figure~\ref{fig1}b-f are in good agreement with our previous study~\cite{LJWang2022}, and we present them here to facilitate the discussion about the links between this quantum bosonic metallic state located in the normal phase and the one appearing close to the superconducting to metallic phase transition tuned by magnetic field in the ultralow temperature limit, making this paper self-contained and complete.

\begin{figure}
\centering
\includegraphics[bb=75 25 373 545,width=5.8cm,height=9.5cm]{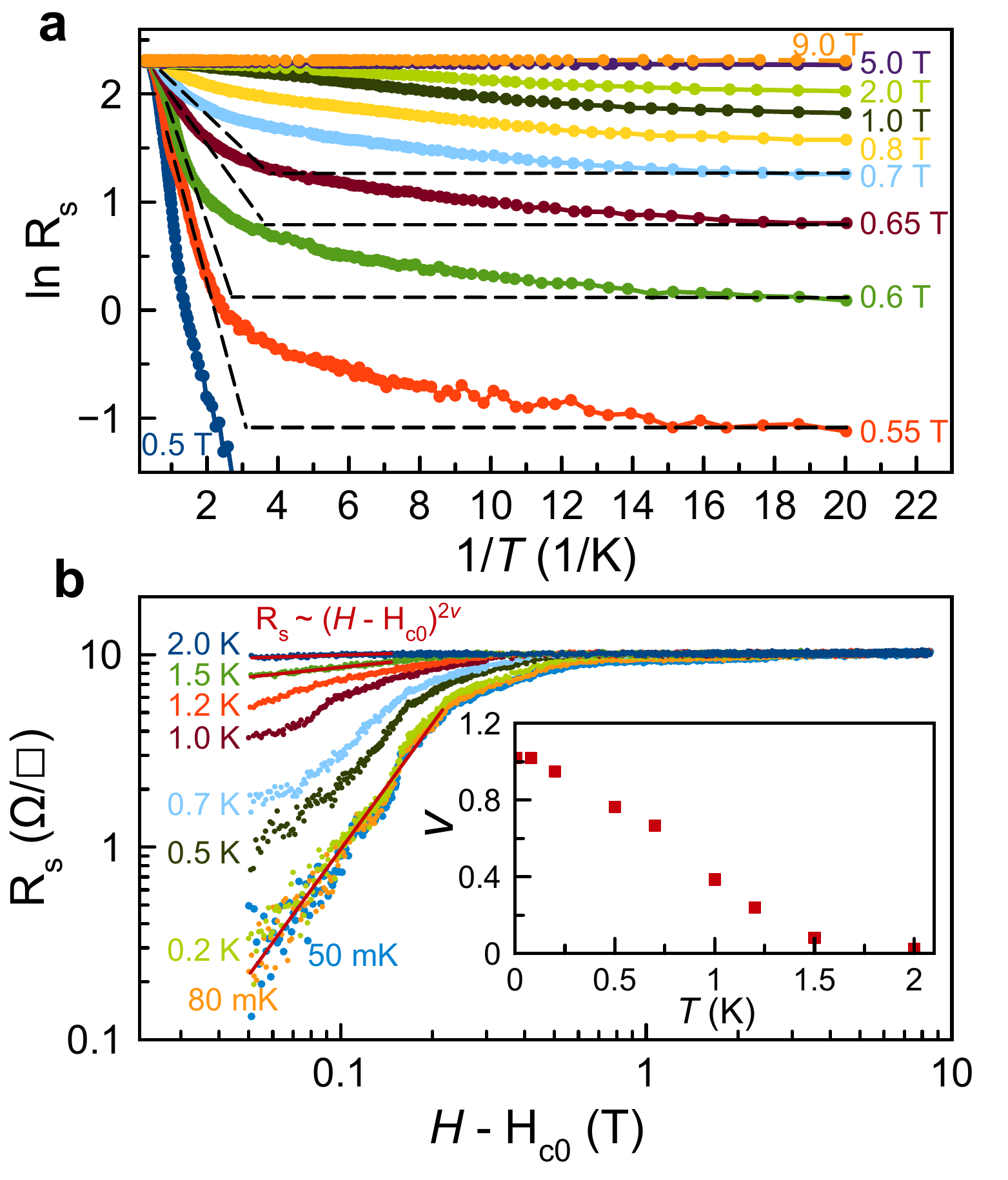}
\caption{Quantum bosonic metallic state in Ti$_2$O$_3$/GaN tuned by field at the ultralow temperatures. (a) Arrhenius plot of the longitudinal electrical resistance $\mathrm{R}_{\mathrm{s}}$ for various magnetic fields perpendicular to the surface of Ti$_2$O$_3$/GaN. In the ultralow temperature regime, the resistance saturates to a finite value, indicating the emergence of a quantum metallic state. (b) Magnetoresistance below the superconducting transition for different temperatures. The data scale to the power-law of Bose-metal $\mathrm{R}_{\mathrm{s}}\sim (H-\mathrm{H}_{\mathrm{c0}})^{2\nu}$. $\nu$ versus $T$ is plotted in the inset.}
\label{fig3}
\end{figure}

To gain further about the underlying nature of the quantum bosonic metallic state in Ti$_2$O$_3$/GaN, a perpendicular magnetic field is applied and the two-dimensional superconductivity-metal transition is investigated. In Figure~\ref{fig2}, we show the temperature-dependent $\mathrm{R}_{\mathrm{s}}$ and the corresponding Hall resistance $\mathrm{R}_{\mathrm{xy}}$ in the presence of an out-of-plane magnetic field. Upon increasing the fields, the superconducting phase is progressively suppressed. If the applied field is $\mu_0H<0.2$ T, a zero-resistance superconducting phase is still present down to $T=1.5$ K (see Figure~\ref{fig2}a), the lowest temperature of this measurement limited. Additionally, we observe a zero $\mathrm{R}_{\mathrm{xy}}$ in this superconducting regime (see Figure~\ref{fig2}b) as expected intuitively because of the electron-hole symmetry of the Bogoliubov quasiparticles that make up the superconducting electron Cooper pairs~\cite{Breznay}. If the field is further increased, a metallic state with a finite $\mathrm{R}_{\mathrm{s}}$ emerges, whereas $\mathrm{R}_{\mathrm{xy}}$ does not seem to change from zero, which is reminiscent of the quantum metallic state at low temperatures. Such a behavior with both zero Hall resistance and finite longitudinal resistance has also been observed previously in disordered two-dimensional superconductors~\cite{Breznay,ZChen2021}. The observation that $\mathrm{R}_{\mathrm{xy}}$ remains negligible at a finite magnetic field suggests that the electron-hole symmetry survives. This apparent electron-hole symmetry behavior heralds the appearance of what has been referred to as the ``elusive'' Bose-metal~\cite{Phillips,Doniach1999}. Notice that we have applied fields as large as 12 T, well above the superconducting critical field $\mu_0\mathrm{H}_{\mathrm{c2}}$, and the superconductivity is eventually evolved into the state that observed in the normal phase (see Figure~\ref{fig2}), which exhibits the striking temperature-independent longitudinal resistance plateau accompanied by the vanished Hall resistance ranging from 1.5 to 15 K that persists in the normal phase (see also Sec. SI in Supporting Information). Considering that the quantum bosonic metallic state located in the normal phase and that proximate to the two-dimensional superconductivity-metal transition tuned by the field in the low temperature regime phenomenologically share these common overall features, we suggest that they possess a unique origin of the Bose-metal.

To further examine the saturating finite resistance in the $T \rightarrow 0 $ K limit, we have performed measurements in the ultralow temperature regime, since it is a fingerprint of the quantum metallic behavior. To get a closer look at the data in this ultralow temperature regime, we arrange data (Figure S4) in the form of an Arrhenius plots, $\ln \mathrm{R}_{\mathrm{s}} \sim 1/T$, as shown in Figure~\ref{fig3}a. As expected, the resistance exhibits a drop with decreasing temperature, followed by a saturating finite resistance, in the zero temperature limit, again a hallmark of the presence of a quantum metallic state~\cite{Nojima2015,Pasupathy2016,TWang}.

In the main panel of Figure~\ref{fig3}b, we also show a log-log plot of magnetoresistance $\mathrm{R}_{\mathrm{s}}$ versus $(H-\mathrm{H}_{\mathrm{c0}})$ taken at various temperatures, where $\mathrm{H}_{\mathrm{c0}}$ is the critical field of the two-dimensional superconductivity-quantum metal transition. Interestingly, the linear scaling behaviors are discernable, implying the existence of a power-law dependence on field. An empirical expression $\mathrm{R}_{\mathrm{s}}\sim (H-\mathrm{H}_{\mathrm{c0}})^{2\nu}$ is used to quantitatively fit the magnetoresistance, and the extracted exponent $\nu$ is plotted in the inset as a function of temperature~\cite{Pasupathy2016,TWang,Doniach2001}. Here, $\nu$ is defined as an exponent of the superfluid correlation length. Remarkably, at ultralow temperatures below 80 mK, magnetoresistances collapse onto a single curve. Results are shown in Figure~\ref{fig3}b ($T = 50$ mK), and yield a critical exponent $\nu=$ 1.02 and a critical field $\mathrm{H}_{\mathrm{c0}}$ = 0.5 T, consistent with the Bose-metal scenario of a quantum metallic state~\cite{Pasupathy2016,TWang,Doniach2001}. These independent and complementary results provide a strong and compelling evidence in support of the intrinsic quantum bosonic metallic state in proximity of the field-induced superconductivity-metal transition, uniquely connected with the one located in the normal phase.

\begin{figure}
\centering
\includegraphics[bb=125 30 395 360,width=5.8cm,height=6.8cm]{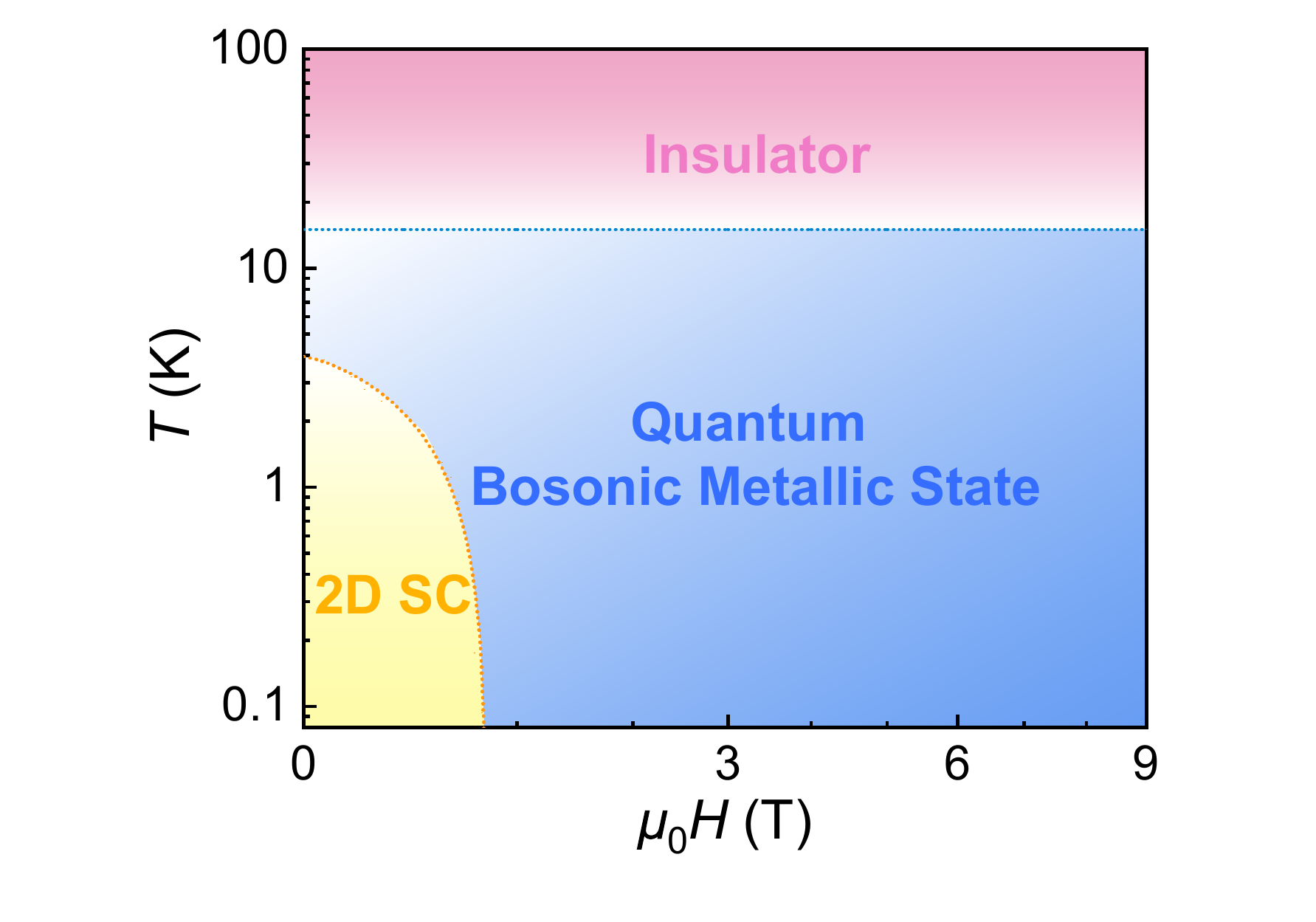}
\caption{Emergence of the quantum bosonic metal in Ti$_2$O$_3$/GaN. Experimental $\mu_0H$-$T$ phase diagram of Ti$_2$O$_3$/GaN, including the insulator, quantum bosonic metallic state, and two-dimensional superconductivity (2D SC). Here, it should be pointed out that the unique quantum bosonic metallic state is experimentally characterized by a temperature-independent plateau of electrical resistance and simultaneously vanishing Hall resistance.}
\label{fig4}
\end{figure}

In Figure~\ref{fig4}, we summarize the intriguing quantum states observed in Ti$_2$O$_3$/GaN in the rich phase diagram of $\mu_0H$-$T$, including the two-dimensional superconductivity and the unique quantum bosonic metallic state. It is interesting to point out that the experimentally observed quantum bosonic metallic state with vanishing Hall resistance inherent to the particle-hole symmetry in Ti$_2$O$_3$/GaN, not only appears in the ultralow temperature limit (see Figure~\ref{fig4}), which is consistent with the previous studies~\cite{Pasupathy2016,TWang,Breznay,ZChen2021}, but also persists in the normal phase (see Figure~\ref{fig4}), which has never been observed in the previous studies~\cite{Pasupathy2016,TWang,Breznay,ZChen2021}. These results suggest the existence of composite bosons formed by electron Cooper pairs that are very robust to survive in the normal phase (see the detailed discussions below and Sec. SI in Supporting Information). This observation shown in Figure~\ref{fig4} is noteworthy, and the intriguing discrepant data by comparison to the previous studies~\cite{Pasupathy2016,TWang,Breznay,ZChen2021} extend our views and ideas of the quantum bosonic metal at superconducting heterointerfaces.

Having experimentally established the existence of a unique quantum bosonic metallic state in Ti$_2$O$_3$/GaN (see the rich phase diagram of Figure~\ref{fig4}), we now delve into the discussions on the underlying key ingredient of pairing nature of heterointerface superconductivity~\cite{Xue2022}. Theoretically, superconductivity is interpreted in terms of composite bosons formed by electron Cooper pairs and their macroscopic long-range phase coherence. Either breaking the electron Cooper pairs or disrupting the long-range phase coherence is expected to be detrimental to superconductivity. When the system is two-dimensional, the long-range phase coherence of electron Cooper pairs dominates the onset of global superconductivity~\cite{Goldman}. In the ultralow temperature regime, the destruction of long-range phase coherence induces a two-dimensional superconductivity-metal transition tuned by an external field. Since electrons usually do not form a conventional metallic state in two dimensions due to the Anderson localization~\cite{Anderson1958,Abrahams1979}, this ultralow temperature transition is theoretically attributed to the dephasing of superconducting electron Cooper pairs based on the bosonic model~\cite{Kapitulnik,Phillips}, in which the electron Cooper pairs persist in this anomalous/quantum metallic state but the long-range phase coherence is lost. As a result, the electron Cooper pairs move diffusively, leading to the metallic-like behavior in the $T\rightarrow 0$ K limit. As Figure~\ref{fig2} and Figures S4 and S5 show, our experimental data behave a vanishing Hall resistance inherent to the electron-hole symmetry, and this is a key ingredient in support of the existence of electron Cooper pairs. Remarkably, these electron Cooper pairs not only are very much alive when the quantum bosonic metallic state emerges in the vicinity of the quantum phase transition at the ultralow temperatures, but also continue to exist in the normal phase (see also the rich phase diagram of Figure~\ref{fig4}). These experimental results thus provide a distinct evidence for electron pairing in the normal phase of Ti$_2$O$_3$/GaN. Besides, it is worth pointing out that a tantalizing puzzle of electron pairing~\cite{Levy2020} at the verge of superconductivity has also been claimed in unconventional~\cite{NPOng2010,HBYang2008,BLKang}, interfacial~\cite{GLCheng2015}, and disordered thin film superconductors~\cite{Ioffe2011,Kopnov2012,Bastiaans}. As a consequence, we mark the ubiquitous composite bosons formed by electron Cooper pairs as a new paradigm for two-dimensional superconductors, which are precursor to the emergence of heterointerface superconductivity.

On the other hand, since the two-dimensional superconductivity and the intriguing quantum bosonic metallic state located at the verge of superconductivity are experimentally observed in Ti$_2$O$_3$/GaN~\cite{LJWang2022} (see also Figure~\ref{fig1}b), while neither of them are perceived in the heterostructure of Ti$_2$O$_3$/non-polar-Al$_2$O$_3$~\cite{LiY2018_1,YLi2018,JXFeng2020}, we thus attribute these emergent quantum states to be the intrinsic nature of interfacial polarization effect induced by the polar GaN substrates, such as interfacial polarons~\cite{Alexandrov1995,Franchini}. Interestingly, these interfacial polarons will interact with each other and give rise to strongly bound bi-polarons with preserving electron-hole symmetry~\cite{Alexandrov1995}, manifesting the behaviors of the quantum bosonic metallic state. In addition, we expect that these bi-polarons could collectively condense into a macroscopically phase-coherent superconducting state eventually when the temperature is decreased. This physical picture of bi-polaronic effects induced by the polar GaN substrates captures the essence of the overall nature of emergent quantum bosonic metallic state and interfacial superconductivity observed in Ti$_2$O$_3$/GaN~\cite{LJWang2022}. Here, it should be worthy noting that this quantum bosonic metallic state behaves in quite a different mechanism compared to the strange/anomalous metallic state with a finite value of transverse Hall resistance and linear-in-temperature longitudinal electrical resistance in cuprates~\cite{YAndo}. Therefore, we suggest a strong candidate of bi-polarons induced by the polar nature of GaN substrates for the physical mechanism of electron pairing in Ti$_2$O$_3$/GaN~\cite{LJWang2022}. Further experiments, including probes of the temperature-dependent tunneling spectra of electronic state evolutions in Ti$_2$O$_3$/GaN, will also be helpful for clarifying the underlying bi-polaronic nature of these appealing quantum states that we observe.

\section*{CONCLUSION}

In summary, we have experimentally uncovered a unique quantum bosonic metallic state phenomenologically characterized by a temperature-independent plateau of electrical resistance and simultaneously vanishing Hall resistance at the superconducting heterointerface of Ti$_2$O$_3$/GaN. This intriguing quantum bosonic metallic state is continuously evolved from the two-dimensional superconductivity, and persists in the normal phase by tuning the magnetic field and/or the temperature, which gradually suppresses the long-range phase coherence of superconducting electron Cooper pairs. Our findings not only elucidate the pivotal clues of understanding of the physics underlying electron pairing mechanism in heterointerface superconductivity, but also stimulate new research upsurge to exploit the emergent intriguing superconducting behaviors at the heterointerfaces and develop next-generation quantum technologies, such as in quantum computing and energy storage.

\section*{EXPERIMENTAL SECTION}

{\bf Thin film growth and structural characterization.} Ti$_2$O$_3$ thin films are grown on GaN(0001) single-crystal substrates by PLD in an ultrahigh vacuum chamber (base pressure of $10^{-9}$ Torr). Prior to growth, the GaN substrates are ultrasonically cleaned with acetone and ethanol. During deposition, a sintered Ti$_2$O$_3$ ceramic target (Kurt J. Lesker Company Inc.) is used to grow the Ti$_2$O$_3$ films with a KrF excimer laser (Coherent 102, wavelength: $\lambda$ = 248 nm). A pulse energy of 110 mJ and a repetition rate of 10 Hz are used. The Ti$_2$O$_3$ films are deposited at a temperature of 750 $^{\circ}$C in a vacuum chamber to promote growth of the superconducting phase. The samples are then cooled to room temperature at a constant rate of 20 $^{\circ}$C/min in a vacuum after deposition. The thickness of the Ti$_2$O$_3$ films is measured by using an AFM (Asylum Research MFP-3D Classic).

{\bf Electrical transport measurements.} The electrical transport measurements are performed using a cryostat with temperature ranging from 1.5 to 300 K (Oxford Instruments TeslatronPT cryostat system), physical properties measurement system with temperature ranging from 50 mK to 4 K (PPMS with dilution refrigerator, Quantum Design), and 10 mK dilution refrigerator (Oxford Instruments Triton 400). The Hall bar structure shown in Figure~\ref{fig1}a and Figure S1 is fabricated by ion-beam etching to systemically measure the electrical transport properties.

\section*{ASSOCIATED CONTENT}

\textbf{Supporting Information}

The Supporting Information is available free of charge at online.

Supporting Text; Optical microscopic image of the fabricated Hall bar devices; Out-of-plane anisotropic superconductivity of Ti$_2$O$_3$/GaN; Temperature-dependent electrical resistance for various applied currents; Electrical transport properties of Ti$_2$O$_3$/GaN at the ultralow temperatures; Ultrahigh magnetic field effects on the observed quantum bosonic metallic state in Ti$_2$O$_3$/GaN.

\section*{AUTHOR INFORMATION}

\noindent\textbf{$^*$Corresponding Author}

\textbf{Gang Mu}, Email: mugang@mail.sim.ac.cn

\textbf{Yan Chen}, Email: yanchen99@fudan.edu.cn

\textbf{Wei Li}, Email: w$\_$li@fudan.edu.cn

\textbf{Notes}

The authors declare no competing financial interest.

\section*{Author Contributions}

G.Z., Y.L., and Z.N. contributed equally to this work. W.L. and Y.C. conceived the project. W.L., C.G., and Y.C. designed the experiments. L.W. grew the samples. G.Z., Y.L., G.L., J.W., Y.W., L.W., Z.N., J.L., J.S., and G.M. performed the electrical transport measurements. H.X. and Z.A. fabricated the Hall bar structure on the thin films. W.L. wrote the paper. All authors gave approval to the final version of the manuscript.

\section*{ACKNOWLEDGMENTS}

This work is supported by the National Natural Science Foundation of China (Grant Nos. 11927807, 92065203, and 12174430), the Strategic Priority Research Program B of Chinese Academy of Sciences (Grant No. XDB33000000), the Beijing Nova Program (Grant No. Z211100002121144), and the Shanghai Science and Technology Committee (Grant Nos. 23ZR1404600 and 20DZ1100604). The authors also thank the Synergetic Extreme Condition User Facility (SECUF) at Institute of Physics, Chinese Academy of Sciences.

\newpage

\begin{figure*}[t!]
\centering
\includegraphics[bb=120 585 280 700,width=6cm,height=4.5cm]{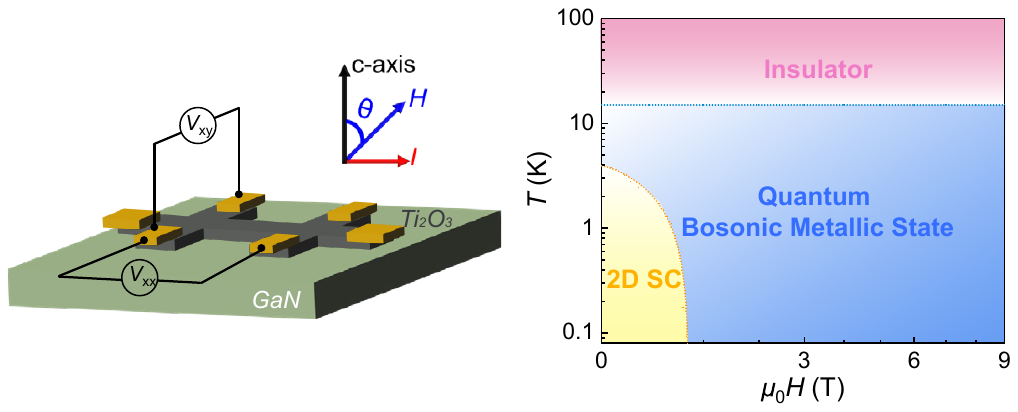}
\caption{\textbf{TOC figure}}
\label{TOC}
\end{figure*}

\end{document}